\documentclass[twocolumn,superscriptaddress,amsmath, amssymb, amsfonts,preprintnumbers,aps,prd,longbibliography,nofootinbib]{revtex4-1}

\usepackage{scalerel}
\usepackage{color}
\usepackage{amsmath,amsfonts,amssymb}%,calrsfs}
\usepackage[small,bf,hang]{caption}
\usepackage{slashed}
\usepackage{latexsym,epsfig}
\usepackage{dsfont}
\usepackage{arydshln}
\usepackage{extarrows}
\usepackage{hyperref}
\usepackage{tikz-cd}
\def\moth{\mathsurround=0pt}
\newdimen\zo \zo=0pt

\def\tick{\leaders\hrule height 0.5ex depth 0pt \hskip 0.5pt}
\def\upboxfill{$\moth \setbox\zo\hbox{\tick}%
  \hskip 3pt\hbox to 0pt{$\tick$\hss}\hrulefill \hbox to 7.5pt{$\tick$\hss}$}

\def\dtick{\leaders\hrule height .34pt depth 0.5ex \hskip 0.5pt}
\def\downboxfill{$\moth \setbox\zo\hbox{\dtick}%
  \hskip 2pt\hbox to 0pt{$\dtick$\hss}\hrulefill \hbox to 2pt{$\dtick$\hss}$}

\def\BV{{\rm BV}_\infty}
\def\cB{{\cal B}}

\def\cO{{\cal O}}

\def\cN{{\cal N}}

\def\cP{{\cal P}}

\def\cV{{\cal V}}

\def\cK{{\cal K}}

\def\cX{{\cal X}}

\def\del{\partial}

\def\B{\square}
\def\Pperp#1{\big\{ #1 \big\}_{\perp}}

\begin{document}

\preprint{
HU-EP-23/15-RTG
}
\title{Weakly constrained double field theory: the quartic theory }

\author{Roberto Bonezzi} 
\email{roberto.bonezzi@physik.hu-berlin.de}
\affiliation{%
Institut f\"ur Physik und IRIS Adlershof, Humboldt-Universit\"at zu Berlin,
Zum Gro{\ss}en Windkanal 2, 12489 Berlin, Germany
}

\author{Christoph Chiaffrino} 
\email{chiaffrc@hu-berlin.de}
\affiliation{%
Institut f\"ur Physik und IRIS Adlershof, Humboldt-Universit\"at zu Berlin,
Zum Gro{\ss}en Windkanal 2, 12489 Berlin, Germany
}

\author{Felipe D\'iaz-Jaramillo} 
\email{felipe.diaz-jaramillo@hu-berlin.de}
\affiliation{%
Institut f\"ur Physik und IRIS Adlershof, Humboldt-Universit\"at zu Berlin,
Zum Gro{\ss}en Windkanal 2, 12489 Berlin, Germany
}

\author{Olaf Hohm} 
\email{ohohm@physik.hu-berlin.de}
\affiliation{%
Institut f\"ur Physik und IRIS Adlershof, Humboldt-Universit\"at zu Berlin,
Zum Gro{\ss}en Windkanal 2, 12489 Berlin, Germany
}

% \date{\today}
\begin{abstract}

Double field theory was originally introduced  as the subsector of closed string field theory on a toroidal background 
given by   the massless fields together with all their massive Kaluza-Klein and winding modes. These massive modes 
are encoded in the dependence of the massless fields on doubled toroidal coordinates, subject to the so-called 
`weak constraint'. This  theory was constructed 
by Hull and Zwiebach in 2009 to cubic order in fields, but due to the weak constraint 
it is a highly non-trivial problem to extend this to quartic and higher order. In this letter we announce and outline  the construction 
of weakly constrained  double field theory to quartic order, in which all coordinates are toroidal and doubled. 
To this end we use the framework of homotopy algebras and  obtain double field theory as a  
double copy of the kinematic homotopy algebra of Yang-Mills theory.

\end{abstract}

\maketitle

\section{Introduction}

One of the core problems of fundamental physics is to construct a workable theory of quantum gravity 
that combines   quantum field theory and general relativity. Among other  research programs on quantum gravity  
string theory arguably plays  a unique role in that it takes the non-renormalizability of general relativity (GR) 
to suggest  that already the \textit{classical} theory will  be replaced by another theory that only at low energies reduces to  
GR. The historic precedent is Fermi's theory of weak interactions, which is non-renormalizable 
and was replaced by the renormalizable electroweak theory featuring new  spin-1 gauge  modes that 
are  rendered massive by the Higgs mechanism \cite{Weinberg:1967tq}. 
At energies much lower than the mass scale of these massive  modes, the latter may be integrated out 
to recover Fermi's theory.

Similarly, one would like to find a renormalizable or perhaps even UV-finite theory for the gravitational field 
but including massive modes such that at low energies one recovers GR. String theory 
provides just that, but it does so at the cost of some  excesses, as 
the inclusion of massive string modes of all possible spins. 
It is hence reasonable to ask whether there might be `smaller' theories of quantum gravity, 
perhaps consistent subsectors of string theory, that inherit some of its unique characteristics such as 
dualities and new effects governed by the parameter $\alpha'$ (of dimension length-square). 
Ideally, such  a theory would be defined by an explicit Lagrangian for  a more 
manageable field content.

A promising subsector is the double field theory (DFT) as originally envisioned  by Hull and Zwiebach \cite{Hull:2009mi}. 
For a toroidal background,  DFT contains the universal massless fields of string theory on 
flat space (the tensor $e_{\mu\bar{\nu}}$, encoding metric  and $B$-field fluctuations, and the dilaton),  
but also  all of their massive Kaluza-Klein and winding modes, which are encoded in a doubled 
coordinate dependence of the massless fields along the toroidal dimensions, thereby exhibiting $O(d,d,\mathbb{Z})$ duality. 
DFT thus contains infinite towers of massive string modes while omitting all other  string modes. 
Such a theory can in principle (though arguably not in practice) be obtained from the full closed string field theory by 
integrating out all modes that do not belong to the DFT sector \cite{Sen:2016qap,Arvanitakis:2020rrk,Arvanitakis:2021ecw}. 
As shown by Sen, this theory would inherit the UV-finiteness of the full string theory \cite{Sen:2016qap}. 
Constructing this theory from the `ground up' is  then a promising path toward 
quantum gravity or at  least the beginning of a `bootstrapping' of quantum gravity for such backgrounds.

Hull and Zwiebach constructed  double field theory in 2009  to cubic order in fluctuations \cite{Hull:2009mi}, but 
it has remained as an unsolved problem to extend this to quartic and ultimately to all orders. 
The theory is subject to the so-called `weak constraint' originating from the level-matching constraint. 
For the construction in \cite{Hull:2009mi} 
this constraint can be implemented in a relatively straightforward manner (see also sec.~3.3 in \cite{Hohm:2022pfi}) 
 but it is a highly non-trivial problem  to extend this to quartic  order.

The subsequent work on DFT has therefore focused on a subsector where fields are subject to the `strong constraint', 
which reduces, in a duality invariant manner, the coordinates to that of the usual 
(undoubled) spacetime \cite{Siegel:1993th,Hull:2009zb,Hohm:2010jy,Hohm:2010pp}. 
Therefore, strongly constrained DFT essentially is a reformulation of GR coupled to $B$-field and dilaton 
(also referred to as `${\cal N}=0$ supergravity'), but it  brings out 
stringy features such as $O(d,d)$ invariance that in conventional supergravity only appear upon further truncations, 
as in dimensional reduction. 
This  has led to powerful applications, for instance in relation to double copy \cite{Hohm:2011dz,Diaz-Jaramillo:2021wtl} and 
in the context of `exceptional field theory', e.g.~\cite{Berman:2012vc,Hohm:2013pua,Hohm:2013vpa,Hohm:2014qga,Malek:2019eaz}, 
which are generalizations  
of DFT in which $O(d,d)$ is extended to the exceptional groups $E_{d+1(d+1)}$ for $d=5,\ldots, 8$.

In this letter we announce and outline the construction of weakly constrained DFT to quartic order. 
The detailed  proofs will appear in \cite{companion}.  
 The consistency of the cubic theory is essentially due to a `kinematical accident', but   to quartic order 
the consistency problems  return with full force. Consequently, if the quartic theory 
exists, it is almost  certain that the  theory exists to all orders.

Our construction uses  two ingredients: the double copy from  scattering amplitudes \cite{Bern:2008qj,Bern:2019prr}, 
which relates gauge theory to gravity amplitudes, 
and the homotopy algebra formulation of field theory \cite{Zwiebach:1992ie,Hohm:2017pnh}. 
We start from  Yang-Mills theory, viewed as a  homotopy Lie or $L_{\infty}$ algebra, and `strip off' color, 
which 
leads to a `kinematic algebra' ${\cal K}$ that is   commutative associative `up to homotopy' \cite{Zeitlin:2008cc}. 
The `color-kinematics duality'  known  from scattering amplitudes suggests that ${\cal K}$ also 
carries a  hidden `Lie'-type algebra. We find a structure that comes close 
to be an $L_{\infty}$ algebra, but it is obstructed 
due to the flat-space  wave operator
$\B=\partial^{\mu}\partial_{\mu}$ being of second order  so that it does not obey  the Leibniz rule 
w.r.t.~the point-wise product of functions. Intriguingly, this is precisely 
the same kind of consistency problem as in weakly constrained DFT.
One may formalize these `$\B$-failures' and  define a more general 
algebraic structure \cite{Reiterer:2019dys}.

For the double copy of Yang-Mills theory one  introduces a second copy $\bar{\cal K}$ of 
the kinematic algebra and seeks  the $L_{\infty}$ algebra of gravity on ${\cal K}\otimes\bar{\cal K}$. 
This space consists of functions of doubled coordinates (coordinates $x$
associated to ${\cal K}$ and coordinates $\bar{x}$ associated to $\bar{\cal K}$), 
but it does not carry an $L_{\infty}$ algebra. Rather, as for the kinematic algebra of Yang-Mills theory, 
this algebra is obstructed by `$\Delta$-failures', where $\Delta=\frac{1}{2} (\B-\bar{\B})$ is the wave operator
on the doubled space of signature $(d,d)$. Thus, in order to obtain an $L_{\infty}$ algebra and hence 
a consistent field theory, one has to impose  $\Delta=0$, thereby `gluing together' 
the two factors by imposing  $\B=\bar{\B}$. Imposing this constraint `strongly', by identifying 
coordinates $x$ with coordinates $\bar{x}$, one obtains the strongly constrained DFT that encodes  
${\cal N}=0$ supergravity. Here we will impose this constraint `weakly', assuming $\B=\bar{\B}$ on functions 
 encoding  genuine winding modes, and we construct  the  $L_{\infty}$ algebra and hence weakly constrained DFT to quartic oder.  
Thereby we  not only solve a major problem in double field theory, but we do so in  a 
way that opens up new   research avenues for  deriving  color-kinematics duality and double copy 
from first principles.

\section{$L_\infty$ algebra of Yang-Mills theory } 

We begin by explaining how Yang-Mills theory 
is encoded in an $L_\infty$ algebra. This algebra is defined on 
a graded vector space, with subspaces in increasing degrees containing  gauge parameters, fields, field equations and Noether identities. The gauge algebra, nonlinear interactions and so on are described by multilinear brackets $B_n$. For instance, the action for 
fields $\psi$ reads 
\begin{equation}\label{firstLinftyAction}
\begin{split}
  S  = \ &\tfrac{1}{2} \big\langle \psi , B_1(\psi)\big\rangle  + \tfrac{1}{3!} \big\langle \psi, B_2(\psi,\psi)\big\rangle \\[0.5ex] 
  &+\tfrac{1}{4!} \big\langle \psi, B_3(\psi,\psi,\psi)\big \rangle +\cdots\;. 
 \end{split} 
 \end{equation} 
The $B_n$ obey quadratic relations (generalized Jacobi identities), which  in turn ensure perturbative consistency of the classical theory.  

In order to display this structure for Yang-Mills theory we begin with the action written  as:
\begin{equation}\label{YMaction}
\begin{split} 
S&=\int d^dx\,\Big[\tfrac12 A^\mu_a\B A_\mu^a-\tfrac12 \varphi_a\varphi^a+\varphi_a \,\del^\mu A_\mu^a\\
&-f_{abc}\,\del_\mu A_\nu^a A^{\mu b}A^{\nu c}-\tfrac14 f^e{}_{ab}f_{ecd} A_\mu^a A_\nu^b A^{\mu c}A^{\nu d}\Big]\;,    
\end{split} 
\end{equation}
where the $\varphi_a$ are auxiliary fields, and $f_{abc}$ are the structure constants of the color Lie algebra $\mathfrak{g}$.
The cubic and quartic vertices are standard, and one recovers the more familiar  action 
after  integrating out $\varphi_a$. 

In Yang-Mills theory every object takes values in the color Lie algebra $\mathfrak{g}$. It is thus natural to view the $L_\infty$ vector space of Yang-Mills theory as the tensor product $\cK\otimes\mathfrak{g}$, where elements of $\cK$ are color-stripped local functions on spacetime. 
The elements of the graded vector space $\cK=\bigoplus_{i=0}^3\cK_i$ can be displayed as:
\begin{equation}\label{Kdiagram}
\begin{tikzcd}[row sep=2mm]
\cK_{0} & \cK_{1} & \cK_2 & \cK_3\\
\lambda& A^\mu & E\\[2mm]
 &\arrow{ul}{b}\varphi&\arrow{ul}{b}E^\mu&\arrow{ul}{b}\cN
\end{tikzcd}\;
\end{equation}
By abuse of language, we refer to $\lambda$ as the gauge parameter, $(A^\mu,\varphi)$ as the fields, 
$(E^\mu,E)$ as the  field equations and $\cN$ as the Noether identity, even though these objects  are \emph{not} Lie algebra-valued. In the above diagram we have also included the action of the $b$ operator, to be discussed below.

The $L_\infty$ algebra  on $\cK\otimes\mathfrak{g}$ is encoded in the Lie algebra  $\frak{g}$ and a 
$C_\infty$ algebra structure on $\cK$ \cite{Zeitlin:2008cc,Borsten:2021hua,Bonezzi:2022yuh}. The latter 
is a homotopy generalization of a commutative associative algebra, where a graded vector space is equipped with multilinear products $m_n$ obeying a set of quadratic relations.
For Yang-Mills theory, the only non-vanishing products are an operator $m_1$ of degree $+1$, a bilinear product $m_2$ of degree zero and a trilinear product $m_3$ of degree $-1$. The nontrivial quadratic relations for  $\psi_i\in\cK$ read
\begin{widetext}
\begin{equation}\label{Coo}
\begin{split}
m_1^2(\psi)&=0\;,\\
m_1\,m_2(\psi_1,\psi_2)&=m_2(m_1\psi_1,\psi_2)+(-1)^{\psi_1}m_2(\psi_1,m_1\psi_2)\;,\\
m_2\big(m_2(\psi_1,\psi_2),\psi_3\big)-m_2\big(\psi_1,m_2(\psi_2,\psi_3)\big)&=m_1\,m_3(\psi_1,\psi_2,\psi_3)+m_3(m_1\psi_1,\psi_2,\psi_3)\\&+(-1)^{\psi_1}m_3(\psi_1,m_1\psi_2,\psi_3)+(-1)^{\psi_1+\psi_2}m_3(\psi_1,\psi_2,m_1\psi_3)\;,   \end{split}    
\end{equation}  
\end{widetext}
expressing that $m_1$ is a nilpotent derivation of $m_2$, and $m_2$ is associative up to homotopy.  
Furthermore, 
the products of a $C_\infty$ algebra have to vanish under shuffle permutations of the arguments. This implies that $m_2$ is graded symmetric: $m_2(\psi_1,\psi_2)=(-1)^{\psi_1\psi_2}m_2(\psi_2,\psi_1)$, while the symmetry of $m_3$ is simpler to express in terms of a redefined product $m_{3h}(\psi_1,\psi_2,\psi_3):=\tfrac13\big(m_3(\psi_1,\psi_2,\psi_3)+(-1)^{\psi_1\psi_2}m_3(\psi_2,\psi_1,\psi_3)\big)$. The original $m_3$ can be reconstructed from $m_{3h}$, so both encode equivalent 
information. 
The product $m_{3h}$  is graded symmetric in the first two arguments and vanishes upon total graded symmetrization. As an example, 
we give the two-product between fields: 
\begin{equation}\label{m2YM} 
\begin{split}
m_2^\mu(A_1,A_2)&=
2\Big( 2\, A_{[1}^\nu\del_\nu A_{2]}^\mu+ \del^\mu A_{[1}^\nu A^{}_{2]\nu} + \del_\nu A^\nu_{[1}\,A_{2]}^\mu\Big) \\
&\equiv (A_1\bullet A_2)^{\mu}  \, ,   
\end{split} 
\end{equation}    
using   the $\bullet$ notation introduced in \cite{Bonezzi:2022yuh}.  

The $L_{\infty}$ algebra of Yang-Mills theory is now defined on the tensor product ${\cal K}\otimes \frak{g}$, 
i.e.,  the objects in (\ref{Kdiagram}) are made $\frak{g}$-valued: 
$x = u^a\otimes t_a$, 
where $t_a$ are the generators of  $\frak{g}$. The 
$B_n$  encoding the $L_{\infty}$ structure   are: 
 \begin{equation}
  \begin{split}
   B_1(x) &= m_1(u^a)\otimes t_a\;, \\
   B_2(x_1,x_2) &= (-1)^{x_1} m_2(u_1^a, u_2^b) f_{ab}{}^{c} \otimes t_c\;,
  \end{split}
 \end{equation} 
with a similar formula for $B_3$ \cite{Bonezzi:2022yuh}.  
In particular, (\ref{m2YM}) and  (\ref{firstLinftyAction}) yield 
the cubic vertex of Yang-Mills theory. 
The $C_{\infty}$ relations (\ref{Coo}), together 
with the Jacobi identities for $f_{abc}$, 
ensure  gauge covariance of the theory.

\section{The $\BV^\B$ kinematic algebra}

Beyond the $C_\infty$ algebra, $\cK$ carries a much richer structure, dubbed $\BV^\B$ in \cite{Reiterer:2019dys}, which plays a crucial role in the double copy construction. The starting point of this new layer is the $b$ operator \cite{Reiterer:2019dys,Ben-Shahar:2021doh,Ben-Shahar:2021zww,Borsten:2022vtg,Bonezzi:2022bse,Borsten:2023reb} whose action on $\cK$ is 
displayed in \eqref{Kdiagram}. 
For instance, $b$ acts on fields $(A^\mu,\varphi)$ by picking out the scalar component $\varphi$ and then 
degree shifting by $-1$ (or, equivalently, viewing $\varphi$ as  a 
gauge parameter). 
It follows that  $b$ is a second differential,  of opposite degree to $m_1$, satisfying the defining  relations 
\begin{equation}\label{bproperties}
 b^2=0\;,\quad b\,m_1+m_1b=\B\;,\quad |b|=-1\;,   
\end{equation}
where $\B=\del^\mu\del_\mu$ is the wave operator or Laplacian, depending on the signature. 
Importantly, so defined $b$  is local and contains no spacetime derivatives.

We next note that $b$ is not compatible with the $C_\infty$ algebra in that 
it does not act as a derivation on the product $m_2$. Rather, its failure to do so 
 defines a new `Lie-type' bracket $b_2$: 
\begin{equation}\label{b2}
\begin{split}
b_2(\psi_1,\psi_2):=\, b m_2(\psi_1,\psi_2)&-m_2(b\,\psi_1,\psi_2)\\
&-(-1)^{\psi_1}m_2(\psi_1,b\,\psi_2)\;.    
\end{split}
\end{equation}
If the bracket $b_2$ and the product $m_2$ are compatible in the graded Poisson sense, \emph{i.e.}~if they obey
\begin{equation}\label{HopePoisson}
\begin{split}
b_2\big(\psi,m_2(\psi_1,\psi_2)\big)&=m_2\big(b_2(\psi,\psi_1),\psi_2\big)\\
&+(-1)^{\psi_1(\psi+1)}m_2\big(\psi_1,b_2(\psi,\psi_2)\big),     
\end{split}    
\end{equation}
then $b_2$ is a graded Lie bracket and the triplet $(b,m_2,b_2)$ forms a BV algebra. In Yang-Mills theory, however, \eqref{HopePoisson} holds only up to homotopy and further `$\B-$deformations', defining a  $\BV^\B$ algebra \cite{Reiterer:2019dys}.

These  $\B-$deformations are a consequence of $\B$ not obeying the Leibniz rule w.r.t.~the point-wise  product of functions, i.e., 
 \begin{equation}
  \B(f\cdot g) - \B f\cdot g -f\cdot \B g = 2\, \partial^{\mu}f\cdot \partial_{\mu}g \;. 
 \end{equation} 
In order to keep track of such  $\B$-failures in the context of bilinear and trilinear maps,  
which is the level studied in \cite{Bonezzi:2022bse}, we find it convenient to define the operators $d_s$ and $d_\B$ as
\begin{widetext}
\begin{equation}\label{ds}
\begin{split}
d_\B(\psi_1,\psi_2)&:=2\,(\del^\mu\psi_1,\del_\mu\psi_2)\;,\\
d_s(\psi_1,\psi_2,\psi_3)&:=2\,(\del^\mu\psi_1,\del_\mu\psi_2,\psi_3)\;,\\
d_\B(\psi_1,\psi_2,\psi_3)&:=2\,(\del^\mu\psi_1,\del_\mu\psi_2,\psi_3)
+2\,(\psi_1,\del^\mu\psi_2,\del_\mu\psi_3)+2\,(\del^\mu\psi_1,\psi_2,\del_\mu\psi_3)\;.    
\end{split}    
\end{equation}
\end{widetext}
For instance, the simplest deformation occurs in the Leibniz relation between the original differential $m_1$ and the bracket $b_2$:
\begin{equation}
\begin{split}
m_1 b_2(\psi_1,\psi_2)+b_2(m_1\psi_1,\psi_2)&+(-1)^{\psi_1}b_2(\psi_1,m_1\psi_2)\\
&=2\,m_2(\del^\mu\psi_1,\del_\mu\psi_2)\,,
\end{split}    
\end{equation}
which can be compactly written as $[m_1,b_2]=m_2d_\B$ in an input-free notation developed in \cite{Bonezzi:2022bse}. In the same notation, the compatibility \eqref{HopePoisson} is relaxed both by a term $[m_1,\theta_3]$, with a trilinear homotopy map $\theta_3$, and by $\B-$deformations governed by $m_{3h}$. Similarly, the Jacobi identity of the bracket $b_2$ holds up to homotopy 
and up to a $\B-$deformation: 
\begin{equation}\label{JacdefK}
3\,b_2b_2\pi+[m_1,b_3]+3\,\theta_3d_s\pi=0\;,    
\end{equation}
where $\pi$ performs graded symmetrization of three arguments.
All relations of the $\BV^\B$ algebra at this order, together with the explicit maps for the Yang-Mills case, are given in \cite{Bonezzi:2022bse}. Let us note  that in the $\BV^\B$ algebra the brackets $(b_1\equiv m_1, b_2,b_3,\ldots)$ form an obstructed $L_\infty$ algebra, with obstructions depending on $\B$. The strategy for constructing DFT as a double copy is to 
eliminate  these obstructions on  a suitable subspace of ${\cal K}\otimes \bar{\cal K}$.

\section{$\BV^\Delta$ algebra on $\cK\otimes\bar\cK$}

Given the $\BV^\B$ algebra on ${\cal K}$, we  expect  that there is a similar structure on the tensor product $\cK\otimes\bar\cK$ of two copies of $\cK$. We present here the resulting $\BV^\Delta$ algebra, which will be the starting point for constructing weakly constrained DFT at quartic order.

Let us then consider the vector space  $\cX:=\cK\otimes\bar\cK$. Since elements of $\cK$ and $\bar\cK$ are functions of coordinates $x^\mu$ and $\bar x^{\bar\mu}$, respectively, elements of $\cX$ are naturally viewed 
as  functions of doubled coordinates $(x^\mu,\bar x^{\bar\mu})$. $\cX$ then comes equipped with two combinations of Laplacians and 
two possible $b-$operators:
\begin{equation}\label{deltabs}
\begin{split}
\Delta_+&:=\tfrac12(\B+\bar\B)\;,\qquad 
\Delta:=\tfrac12(\B-\bar\B)\;,\\[2mm]
b^\pm&:=\tfrac12\big(b\otimes\mathds{1}\pm\mathds{1}\otimes\bar b\big)\;,
\end{split}    
\end{equation}
where the notation shows that the $b$ and $\bar b$ operators of $\cK$ and $\bar\cK$ act on the respective factors of the tensor product.
The 
vector space carrying the $L_\infty$ algebra of weakly constrained DFT, which we denote $\cV_{\rm DFT}$, is the linear subspace of  $\cX$
annihilated by both $\Delta$ and $b^-$ \cite{Hull:2009mi,Bonezzi:2022yuh}:
\begin{equation}\label{VDFT}
\cV_{\rm DFT}:={\rm ker}\Delta\,\cap\,{\rm ker}b^-\subset\cX\;.    
\end{equation}

We now start to present the relevant algebraic structures on $\cX$. The proofs and details of this construction will be given in \cite{companion}. A nilpotent differential and  two-product can be  defined by
\begin{equation}
\begin{split}
M_1&:=m_1\otimes\mathds{1}+\mathds{1}\otimes\bar m_1\;,\\
M_2&:=m_2\otimes\bar m_2\;.
\end{split}
\end{equation}
Using  that $m_1$ and $\bar m_1$ are derivations of the respective two-products, c.f.~\eqref{Coo}, it quickly follow 
that $M_1$ is a derivation of $M_2$:
\begin{equation}
[M_1,M_2]=0 \,,    
\end{equation}
using the input-free notation of \cite{Bonezzi:2022bse}. One can prove that $M_2$ is associative up to homotopy, with a three-product of the schematic form $M_{3h}=\frac{1}{2}(m_{3h}\otimes\bar m_2\bar m_2+\cdots)$. 
$M_1$, $M_2$ and $M_{3h}$ start forming a $C_\infty$ algebra on $\cX$ which we  expect  to have 
infinitely many higher products. 

This $C_\infty$ algebra on $\cX$ is part of yet larger `BV-type' algebra. To see this, 
consider  the  $b^-$ operator in \eqref{deltabs}, which obeys
\begin{equation}\label{M1Delta}
[M_1,b^-]=\Delta\;.  
\end{equation}
This is the beginning of  $\BV^\Delta$ algebra on $\cX$, which obeys the same relations as the $\BV^\B$ algebra   on $\cK$, 
upon replacing $b\rightarrow b^-$, $\B\rightarrow\Delta$, $m_n\rightarrow M_n$, etc. In particular, the $\Delta-$deformations act, up to three inputs, as 
\begin{equation}
D_s:=\tfrac12\,(d_s-\bar d_s)\;,\quad D_\Delta:=\tfrac12\,(d_\B-\bar d_{\bar\B})\;,    
\end{equation}
with $\bar d_s$, $\bar d_{\bar\B}$ as in \eqref{ds} but in terms of barred derivatives. 
The two-bracket $B_2$ is defined analogously to \eqref{b2}: 
\begin{equation}
\begin{split}
B_2&:=[b^-, M_2]=\tfrac12\,\big(b_2\otimes\bar m_2-m_2\otimes\bar b_2\big)\;, 
\end{split}    
\end{equation}
with the second form manifesting  the Lie $\otimes$ Associative structure of the tensor product. As for $b_2$ on $\cK$, $M_1$
fails to be a derivation of $B_2$ by a $\Delta-$obstruction:
\begin{equation}\label{Leibnizdelta}
[M_1,B_2]=M_2D_{\Delta}  \;, 
\end{equation}
and the Poisson compatibility \eqref{HopePoisson} between $B_2$ and $M_2$ is deformed in the same manner, with an homotopy $\Theta_3$ and $\Delta-$deformations proportional to $M_{3h}$. The Poisson homotopy $\Theta_3$ can be determined from the underlying Yang-Mills maps and has the  form
\begin{equation}
\begin{split}
\Theta_3&=\Theta_{3s}-[b^-,M_{3h}]\;,
\end{split}    
\end{equation}
where $\Theta_{3s}$ is the graded symmetric part of $\Theta_{3}$
of the schematic form $\Theta_{3s}=\tfrac12  \theta_{3s}\otimes\bar m_2\bar m_2\Pi+\cdots $, 
where $\Pi$ enforced graded symmetrization.  
The $\Theta_3$ map governs the $\Delta-$deformation of the Jacobi identity of $B_2$:
\begin{equation}
3\,B_2B_2\Pi+[M_1,B_3]+3\,\Theta_{3}\,D_s\,\Pi=0\;,    
\end{equation}
and determines the three-bracket as $B_3=-[b^-,\Theta_{3s}]$.

\section{Weakly constrained DFT}

In order to transport the algebraic structure of $\cX$ to the DFT subspace $\cV_{\rm DFT}$ we shall proceed in two steps: 
we will first employ a notion called homotopy transfer to define a $\BV^\Delta$ algebra on ${\rm ker}\,\Delta$, and then restrict further to 
${\rm ker}\,b^-$, where only the $L_\infty$ subsector survives.

Homotopy transfer is defined by a projection map from a larger to a smaller space, 
so that the so-called homotopy relation is obeyed \cite{Arvanitakis:2020rrk,Arvanitakis:2021ecw,Erbin:2020eyc,Koyama:2020qfb}. 
The projector  $\cP_\Delta$ onto 
 ${\rm ker}\,\Delta$ will be  defined for the special case that the  background geometry is a doubled Euclidean torus, with square doubled metric $(\delta_{\mu\nu}, \delta_{\bar\mu\bar\nu})$. Functions  on the doubled torus can be expanded in Fourier modes:
\begin{equation}\label{Fourier}
f(x,\bar x)=\sum_{k,\bar k}\tilde f(k,\bar k)\,e^{ik\cdot x+i\bar k\cdot\bar x}\;,    
\end{equation}
with integer momenta $(k^\mu, \bar k^{\bar\mu})\in\mathbb{Z}^{2d}$. Since $\Delta$ acts as $-\frac12 (k^2-\bar k^2)$ on the modes, the projector is explicitly defined as
\begin{equation}
\big(\cP_\Delta f\big)(x,\bar x)=\sum_{k,\bar k}\delta_{k^2-\bar k^2, 0}\,\tilde f(k,\bar k)\,e^{ik\cdot x+i\bar k\cdot\bar x}\;,    
\end{equation}
and obeys $\cP_\Delta\Delta=\Delta\cP_\Delta=0$. Next, we need to find a homotopy map $h$, which is a degree $-1$ operator obeying
\begin{equation}\label{homotopy relation}
\big[M_1,h\big]=1-\cP_\Delta \;. 
\end{equation} 
Thanks to $b^-$ obeying \eqref{M1Delta}, such an homotopy can be defined as
\begin{equation}\label{h}
h:=b^-G\,(1-\cP_\Delta)\;,    
\end{equation}
where $G$ is the $\Delta-$propagator, acting as $-{2}(k^2-\bar k^2)^{-1}$ on the Fourier modes, which is the inverse of $\Delta$ away from its kernel. The homotopy transferred maps, which we denote with an overline, can be computed systematically once \eqref{homotopy relation} is given. For instance, the transferred brackets $\overline{B}_n$ are given by
\begin{equation}\label{Bbars}
\begin{split}
\overline{B}_1&={M}_1\big\rvert_{{\rm ker}\Delta}\;,\\[0.5ex] 
 \overline{B}_2&=\cP_\Delta\, B_2\big\rvert_{{\rm ker}\Delta}\;,\\[0.5ex]
\overline{B}_3&=\cP_\Delta\,\Big(B_3-3\,B_2hB_2\,\Pi\Big)\Big\rvert_{{\rm ker}\Delta}\;,    
\end{split}    
\end{equation}
with $\rvert_{{\rm ker}\Delta}$ denoting restriction of the inputs to ${\rm ker}\Delta$.

Let us emphasize   that the $\BV^\Delta$ algebra is \textit{not} transported to an undeformed $\BV$ on ${\rm ker}\,\Delta$. 
While the obstructions of the form $D_\Delta$, being $\Delta-$commutators, vanish on ${\rm ker}\Delta$, 
the $D_s$ obstructions do not. As a consequence, while the transferred differential $\overline{B}_1$ is now a derivation of $\overline{B}_2$,  \emph{i.e.}~$[\overline{B}_1,\overline{B}_2]=0$, ensuring a consistent cubic theory, the homotopy Jacobi identity of $\overline{B}_2$ is still obstructed:
\begin{equation}
3\,\overline{B}_2\overline{B}_2\,\Pi+[\overline{B}_1,\overline{B}_3]+3\,\overline{\Theta}_{3h}\,D_s\,\Pi=0\;,
\end{equation}
where $\overline{\Theta}_{3h}=\overline{\Theta}_{3}-\overline{\Theta}_{3s}$. 
The obstruction is, however, quite milder than the original one, since the symmetric part of $\overline{\Theta}_3$ does not contribute.

The final step to reach $\cV_{\rm DFT}$ is implemented by
 restricting the inputs $\Psi_i\in{\rm ker}\,b^-$. Among the maps of $\BV^\Delta$, only the $L_\infty$ brackets restrict to ${\ker b^-}$, since they are all given by $b^--$commutators, and $b^-$ is nilpotent. We denote these brackets by $\cB_n$:
\begin{equation}\label{calB12}
\begin{split}
\cB_1&:=\overline{B}_1\rvert_{{\rm ker}b^-}\;,\\
\cB_2&:=\overline{B}_2\rvert_{{\rm ker}b^-}=b^-\overline{M}_2\rvert_{{\rm ker}b^-}\;.
\end{split}    
\end{equation}
The homotopy Jacobi identity of $\cB_2$ is still obstructed:
\begin{equation}\label{Looobstructed}
3\,\cB_2\cB_2\,\Pi+\big[\cB_1,\overline{B}_3\rvert_{{\rm ker}b^-}\big]=\cO\;,   
\end{equation}
where   the obstruction $\cO$ is given by 
\begin{equation}\label{deltaJacfinal}
\cO=3\,\cP_\Delta\,\Big(b^-{M}_{3h}\,D_s-B_2\,(1-\cP_\Delta)\,B_2\Big)\,\Pi\Big\rvert_{\cV_{\rm DFT}}\;,  
\end{equation}
with  the maps on $\cX$ restricted  to ${\cV_{\rm DFT}}={{\rm ker}\Delta\cap{\rm ker}b^-}$.
The only possibility to have an unobstructed $L_\infty$ algebra to this order is that the obstruction $\cO$ is $\cB_1-$exact, 
for $B_3$ can then be redefined. We first note that $\cO$ is $\cB_1-$closed, $[\cB_1,\cO]=0$, 
as a consequence of  \eqref{Looobstructed}. To see that it is also exact, we shall consider the operator $b^+$ in \eqref{deltabs}, which obeys
\begin{equation}
[\cB_1,b^+]=\Delta_+= \tfrac12(\B+\bar\B)
\;.    
\end{equation}
On a doubled Euclidean torus, $\Delta_+$ is negative semi-definite, so that it is invertible except on zero-modes $(k,\bar k )=(0,0)$. However, one may verify  that the obstruction \eqref{deltaJacfinal} does not contain zero-modes, due to the projection structure of the form $\cP_\Delta f\,(1-\cP_\Delta)g$, for functions $f$ and $g$. [For the first term in  \eqref{deltaJacfinal} one needs to use that 
for three weakly constrained functions $D_s(\psi_1\psi_2\psi_3)=\Delta(\psi_1\psi_2) \psi_3=(1-\cP_\Delta)\Delta(\psi_1\psi_2)
\cP_\Delta\psi_3$.]

One can thus invert the Laplacian $\Delta_+$ on $\cO$ and write
\begin{equation}
\cO=\Big[\cB_1,\frac{b^+}{\Delta_+}\Big]\cO=\Big[\cB_1,\frac{b^+}{\Delta_+}\cO\Big]\;,    
\end{equation}
where we used $[\cB_1,\cO]=0$. Since $\cO$ is exact, we can shift $\overline{B}_3\big\rvert_{{\rm ker}b^-}$ appearing in \eqref{Looobstructed} by $\frac{b^+}{\Delta_+}\cO$ and obtain a genuine $L_\infty$ relation on $\cV_{\rm DFT}$:
\begin{equation}
3\,\cB_2\cB_2\,\Pi+[\cB_1,\cB_3]=0\;,   
\end{equation}
where $\cB_1$ and $\cB_2$ are given by \eqref{calB12}, and the final three-bracket reads
\begin{equation}\label{finalB3}
\cB_3=\cP_\Delta\,B_3\big\rvert_{\cV_{\rm DFT}}-\frac{b^+}{\Delta_+}\cO\;,
\end{equation}
with $\cO$ given by \eqref{deltaJacfinal}. Notice that $B_2hB_2$ in the definition \eqref{Bbars} of the transported $\overline{B}_3$ drops out 
on ${\rm ker}b^-$, due to $h\propto b^-$ and ${B}_2\propto b^-$.

We have thus succeeded in constructing the three-bracket of weakly constrained DFT on a torus. 
In this we introduced a new non-locality due to $\frac{1}{\Delta_+}$ that is, however, perfectly well-defined on the torus. 
One may ask whether this non-locality is removable, which would happen if ${\cal O}=\Delta_+{\cal R}$ for 
some local expression ${\cal R}$. We will show in the next section that this is not possible, by 
computing the explicit 3-brackets for the gauge algebra.

\section{Fields and gauge algebra of DFT}

We begin by reviewing  the DFT vector space ${\cal V}_{\rm DFT}$. 
The gauge parameter $\Lambda=(-\lambda_\mu,\bar\lambda_{\bar\mu}, \eta)$ in degree $-1$ consists of two vector parameters, encoding diffeomorphisms and $B-$field transformations, and a scalar St\"uckelberg parameter. The fields $\psi$ in degree zero comprise the tensor fluctuation $e_{\mu\bar\nu}$, two auxiliary vectors $(f_\mu,\bar f_{\bar\mu})$ and two scalars $(e,\bar e)$, one combination 
of which is the dilaton and the other pure gauge. This coincides  with the original formulation  of \cite{Hull:2009mi}. 
The linear 
gauge transformations $\delta\psi={\cal B}_1(\Lambda)$ yield, in particular, 
$\delta e_{\mu\bar{\nu}}=\partial_{\mu}\bar{\lambda}_{\bar{\nu}}+ \bar{\partial}_{\bar{\nu}}\lambda_{\mu}$, exhibiting the 
double copy structure of linearized diffeomorphisms and $B-$field gauge transformations. 

Our goal now is to  compute the  subsector of the gauge algebra of weakly constrained DFT, restricting for simplicity to 
gauge parameters of the form $\Lambda=(-\lambda_{\mu},0,0)$. The gauge algebra is encoded in the homotopy Jacobi relation 
\begin{equation}\label{HomJacLam}
    \text{Jac}(\Lambda_{1},\Lambda_{2},\Lambda_{2})+[\mathcal{B}_{1}, \mathcal{B}_{3}](\Lambda_{1},\Lambda_{2},\Lambda_{3})=0\; ,
\end{equation}
where the  Jacobiator $\text{Jac}(\Lambda_{1},\Lambda_{2},\Lambda_{3})$ is defined as
\begin{equation}
\begin{split}
    \text{Jac}(\Lambda_{1},\Lambda_{2},\Lambda_{3})
    \stackrel{[123]}{:=} &3\, \mathcal{B}_{2}(\mathcal{B}_{2}(\Lambda_{1},\Lambda_{2}),\Lambda_{3})\; , 
\end{split}
\end{equation}
with the notation indicating complete anti-symmetrization over $123$. 
The relation \eqref{HomJacLam} takes values in the space of gauge parameters and hence has three components. Here  we  will 
work out the component corresponding to $\lambda_{\mu}$, for which the Jacobiator reads
\begin{widetext}
\begin{equation}\label{jacobo2}
\begin{split}
        \text{Jac}(\Lambda_{1},\Lambda_{2},\Lambda_{3})^{\mu}\stackrel{[123]}{=}\tfrac{3}{2}\, \mathcal{P}_{\Delta}\,  &\Big[\del^{\mu}(\lambda_{1\, \rho}\, \del^{\rho}\lambda_{2\, \nu}\, \lambda^{\nu}_{3})+2\, \del_{\rho}\lambda_{1\, \nu}\, \lambda_{2}^{\nu}\, \del^{\rho}\lambda_{3}^{\mu}+\Delta_{+} \lambda_{1\, \rho}\, \lambda^{\rho}_{2}\, \lambda_{3}^{\mu}\\
       &+ 2\, \del_{\rho}\lambda_{1}^{\rho}\, \lambda_{2}^{\nu}\, \del_{\nu}\lambda_{3}^{\mu} +\del_{\rho}\lambda_{1}^{\rho}\,\del^{\mu}\lambda_{2\, \nu}\, \lambda_{3}^{\nu}+\lambda_{2}^{\nu}\, \del_{\nu}\del_{\rho}\lambda^{\rho}_{1}\, \lambda_{3}^{\mu}\Big]\\
       -\tfrac{3}{4}\, \mathcal{P}_{\Delta}\,  &\Big[\Pperp{\lambda_{1}\bullet \lambda_{2}}\bullet \lambda_{3}\Big]^{\mu}\; \in \mathcal{V}_{-1}\; ,
\end{split}
\end{equation}
\end{widetext}
where we used the $\bullet$ notation introduced in (\ref{m2YM}), $\mathcal{P}_{\Delta}\, \B=\mathcal{P}_{\Delta}\, \Delta_{+}$, 
which follows from (\ref{deltabs}),  and we introduced the orthogonal  projector  $\Pperp{\cdot }\equiv 1-\mathcal{P}_{\Delta}$.  

In order to verify  the homotopy Jacobi relation we need the 
following components of $\mathcal{B}_{3}$: first, $\mathcal{B}_{3}$ evaluated on three gauge parameters: 
\begin{equation}
\begin{split}
    \cB_{3}(\Lambda_{1},\Lambda_{2},\Lambda_{3})
    \stackrel{[123]}{=}\tfrac{3}{2}\, \mathcal{P}_{\Delta}\, \big[ \lambda_{1\, \rho}\, \del^{\rho}\lambda_{2\, \nu}\, \lambda_{3}^{\nu} \big] \in \mathcal{V}_{-2}\; , 
\end{split}
\end{equation}
from which one obtains  the $\mu$-component of the action of the differential on $\cB_{3}(\Lambda_{1},\Lambda_{2},\Lambda_{3})$: 
\begin{equation}\label{delB3}
\begin{split}
    \cB_{1}\mathcal{B}_{3}(\Lambda_{1},\Lambda_{2},\Lambda_{3})^{\mu}\stackrel{[123]}{=}-\tfrac{3}{2}\, \mathcal{P}_{\Delta}\, \del^{\mu}\big( \lambda_{1\, \rho}\, \del^{\rho}\lambda_{2\, \nu}\, \lambda_{3}^{\nu} \big)%\; \in \mathcal{V}_{-1}
    \; .
\end{split}
\end{equation}
Second, we have to compute  $ 3\, \mathcal{B}_{3}(\Lambda_{[1},\Lambda_{2},\cB_{1}(\Lambda_{3]}))$, 
for which we need $\cB_{3}$ evaluated on two gauge parameters and a field: 
\begin{widetext}
\begin{equation}\label{B32gauge1field}
\begin{split}
\mathcal{B}_{3}(\Lambda_{1},\Lambda_{2},\Psi)^{\mu}\stackrel{[12]}{=}&\tfrac{1}{2}\, \mathcal{P}_{\Delta} \, \Big[ 2\, f_{\rho}\, \lambda_{1}^{\rho}\, \lambda_{2}^{\mu}+4e\, \lambda_{1}^{\nu}\, \del_{\nu}\lambda_{2}^{\mu}+2\lambda_{1}^{\nu}\, \del_{\nu}e\, \lambda^{\mu}_{2}+2e\, \del^{\mu}\lambda_{1\, \nu}\lambda_{2}^{\nu}\\
&-\, e^{\mu\bar\nu}\, \bar\del_{\bar\nu}\lambda_{1\, \rho}\, \lambda_{2}^{\rho}+\, \bar\del_{\bar\nu}\lambda_{1}^{\mu}\, e^{\rho\bar\nu}\, \lambda_{2\, \rho}-\big( 2\, f^{\rho}+\bar\del_{\bar\nu} e^{\rho\bar\nu} \big)\, \lambda_{1\, \rho}\, \lambda_{2}^{\mu}\Big]\\
&-\mathcal{P}_{\Delta}\, \tfrac{1}{\Delta_{+}}\, \bar\del^{\bar\nu}\Big[ \del_{\rho}\lambda_{1}^{\mu}\, \del^{\rho}\lambda_{2}^{\nu}\, e_{\nu\bar\nu}+\lambda^{\nu}_{1}\, \del_{\rho}\lambda_{2}^{\mu}\, \del^{\rho}e_{\nu\bar\nu}+\del_{\rho}e^{\mu}{}_{\bar\nu}\, \del^{\rho}\lambda_{1\, \nu}\, \lambda_{2}^{\nu} \\
&-\bar \del_{\bar\rho}\lambda_{1}^{\mu}\, \bar\del^{\bar\rho}\lambda_{2}^{\nu}\, e_{\nu\bar\nu}-\lambda^{\nu}_{1}\, \bar\del_{\bar\rho}\lambda_{2}^{\mu}\, \bar\del^{\bar\rho}e_{\nu\bar\nu}-\bar\del_{\bar\rho}e^{\mu}{}_{\bar\nu}\, \bar\del^{\bar\rho}\lambda_{1\, \nu}\, \lambda_{2}^{\nu}\Big]\\
&+\tfrac{1}{4}\, \mathcal{P}_{\Delta}\, \tfrac{1}{\Delta_{+}}\,  \bar\del^{\bar\nu}\big[\Pperp{\lambda_{1}\bullet \lambda_{2}}\bullet e_{\bar\nu}+2\, \lambda_{2}\bullet \Pperp{\lambda_{1}\bullet e_{\bar \nu}}\big]^{\mu}\; \in \mathcal{V}_{-1}\; .
\end{split}
\end{equation}
\end{widetext}
From this expression one infers, e.g.~by inspection of the last  line, that the non-locality inherent in  $\tfrac{1}{\Delta_{+}}$ is 
unavoidable: there is no overall $\Delta_+$ that can be factored out to cancel it, as $\bar\del^{\bar\nu}$ is contracted 
with $e_{\mu\bar{\nu}}$ and not with a derivative. This changes after 
replacing the field  in \eqref{B32gauge1field} by ${\cal B}_1(\Lambda)$.  
For instance, in  the last line in equation \eqref{B32gauge1field} one obtains 
\begin{equation}
\begin{split}
    & \mathcal{P}_{\Delta}\, \tfrac{1}{\Delta_{+}} \bar\del^{\bar\nu}\big[\Pperp{\lambda_{[1}\bullet \lambda_{2}}\bullet \bar\del_{\bar\nu}\lambda_{3]}+2\, \lambda_{[2}\bullet \Pperp{\lambda_{1}\bullet \bar\del_{\bar\nu}\lambda_{3]}}\big]^{\mu} \\ 
  & \ = \    \mathcal{P}_{\Delta}\, \tfrac{1}{\Delta_{+}}\, \bar\del^{\bar\nu}\bar \del_{\bar\nu}\big[\Pperp{\lambda_{[1}\bullet \lambda_{2}}\bullet \lambda_{3]}\big]^{\mu}\; ,
\end{split} 
\end{equation}
where the equality follows using the Leibniz rule and  the antisymmetry of the labels. 
Under the projector $\mathcal{P}_{\Delta}$ we can then use 
the weak constraint $\bar\del_{\bar\nu}\bar\del^{\bar\nu}\equiv \bar\square=\square$ together with $\mathcal{P}_{\Delta}\B=\mathcal{P}_{\Delta}\Delta_{+}$ to cancel $\tfrac{1}{\Delta_{+}}$.  
Performing similar manipulations for the other terms one finally obtains 
\begin{widetext}
\begin{equation}\label{B3del}
\begin{split}
    3\, \mathcal{B}_{3}(\Lambda_{[1},\Lambda_{2},\cB_{1}(\Lambda_{3]}))^{\mu}\stackrel{[123]}{=}-&\tfrac{3}{2}\, \mathcal{P}_{\Delta}\,\Big[ \Delta_{+} \lambda_{3\, \rho}\, \lambda_{1}^{\rho}\, \lambda_{2}^{\mu}
    +2\, \del_{\rho}\lambda_{3}^{\rho}\, \lambda^{\nu}_{1}\, \del_{\nu}\lambda_{2}^{\mu}+\lambda_{1}^{\nu}\, \del_{\nu}\del_{\rho}\lambda_{3}^{\rho}\, \lambda_{2}^{\mu}\\
    +&\del_{\rho}\lambda_{3}^{\rho}\, \del^{\mu}\lambda_{1\, \nu}\, \lambda_{2}^{\nu}
    +2\, \del_{\rho}\lambda_{1}^{\nu}\, \lambda_{2\, \nu}\, \del^{\rho}\lambda_{3}^{\mu}\Big]\\
    +&\tfrac{3}{4}\, \mathcal{P}_{\Delta}\, \big[ \Pperp{\lambda_{1}\bullet \lambda_{2}}\bullet \lambda_{3} \big]^{\mu}\; \in \mathcal{V}_{-1}\; ,
\end{split}
\end{equation}
\end{widetext}
in which the only non-locality is due to the overall $\mathcal{P}_{\Delta}$. 
Finally, adding up \eqref{jacobo2}, \eqref{delB3} and \eqref{B3del}  verifies  the 
homotopy Jacobi relation \eqref{HomJacLam}.

\section{Outlook}

We constructed a weakly constrained double field theory to quartic order with all dimensions toroidal and doubled. 
This remains to be generalized  to Lorentzian backgrounds, with a certain number of 
`external' non-compact and hence undoubled dimensions, and to cosmological backgrounds  \cite{Hohm:2022pfi}. 
Furthermore, the construction needs to be extended to all 
orders, which requires  an   improved understanding of the kinematic algebra of Yang-Mills theory to all orders. 
Finally, it would be interesting to extend this to exceptional field theory and genuine M-theory modes.

\subsection*{Acknowledgements}

We are indebted to Barton Zwiebach for discussions. R.~B. and O.~H. thank MIT for hospitality 
during the final stages of this project.

This work is supported by the Deutsche Forschungsgemeinschaft (DFG, German Research Foundation) - Projektnummer 417533893/GRK2575 ``Rethinking Quantum Field Theory" 
and by 
the European Research Council (ERC) under the European Union's Horizon 2020 research and innovation programme (grant agreement No 771862). 

%\bibliography{WeakDC}


\end{document}